\documentclass[useAMS,usenatbib]{mn2e}

\usepackage{graphicx}
\usepackage{multirow}
\newcommand{\MS}{\ifmmode{\,}\else\thinspace\fi{\rm M}\ifmmode_{\odot}\else$_{\odot}$\fi}
\newcommand{\LS}{\ifmmode{\,}\else\thinspace\fi{\rm L}\ifmmode_{\odot}\else$_{\odot}$\fi}
\newcommand{\Ke}{\ifmmode{\,}\else\thinspace\fi{\rm K}}
\newcommand{\teffs}{\ifmmode T_{\rm eff}^{\rm s}\else$T_{\rm eff}^{\rm s}$\fi}

\title[Testing the Stothers model]{Variable turbulent convection as the cause of the Blazhko effect -- testing the Stothers model}
\author[R. Smolec, P. Moskalik, K. Kolenberg, et al.]{R. Smolec$^{1}$\thanks{E-mail:
radek.smolec@univie.ac.at}, P. Moskalik$^{2}$, K. Kolenberg$^{1}$, S. Bryson$^{3}$, M. T. Cote$^{3}$ and R. L. Morris$^{4}$\\
$^{1}$Institute for Astronomy (IfA), University of Vienna, T\"urkenschanzstrasse 17, A-1180 Vienna, Austria\\
$^{2}$Copernicus Astronomical Centre, Bartycka 18, 00-716 Warszawa, Poland\\
$^{3}$NASA Ames Research Centre, Moffett Field, CA 94035, USA\\
$^{4}$SETI Institute/NASA Ames Research Centre, Moffett Field, CA 94035, USA}
\begin{document}

\date{Accepted . Received ; in original form }

\pagerange{\pageref{firstpage}--\pageref{lastpage}} \pubyear{2010}

\maketitle

\label{firstpage}

\begin{abstract}

The amplitude and phase modulation observed in a significant fraction of the RR~Lyrae variables -- the Blazhko effect -- represents a long-standing enigma in stellar pulsation theory.
No satisfactory explanation for the Blazhko effect has been proposed so far. In this paper we focus on the \cite{st06} idea, in which modulation is caused by changes in the structure of the outer convective zone, caused by a quasi-periodically changing magnetic field. However, up to this date no quantitative estimates were made to investigate whether such a mechanism can be operational and whether it is capable of reproducing the light variation we observe in Blazhko variables. We address the latter problem. We use a simplified model, in which the variation of turbulent convection is introduced into the non-linear hydrodynamic models in an {\it ad hoc} way, neglecting interaction with the magnetic field. We study the light curve variation through the modulation cycle and properties of the resulting frequency spectra. Our results are compared with {\it Kepler} observations of RR~Lyr. We find that reproducing the light curve variation, as is observed in RR~Lyr, requires a huge modulation of the mixing length, of the order of $\pm$50 per cent, on a relatively short time-scale of less than 40\thinspace days. Even then, we are not able to reproduce neither all the observed relations between modulation components present in the frequency spectrum, nor the relations between Fourier parameters describing the shape of the instantaneous light curves.
\end{abstract}

\begin{keywords}
convection -- hydrodynamics -- methods: numerical -- stars: oscillations -- stars: variables: RR~Lyrae -- stars: individual: RR~Lyr
\end{keywords}

\section{Introduction}

The periodic amplitude and phase modulation observed in many RR~Lyrae variables -- the Blazhko effect -- is one of the most important, still unsolved problems in classical pulsation theory. In the last few years, thanks to extensive ground-based observation campaigns \citep[e.g.,][]{kg07,jj09} and satellite missions, {\it CoRoT} \citep{cea10} and {\it Kepler} \citep[e.g.,][]{bea10}, we finally obtained nearly continuous data, allowing for detailed study of the light variation associated with the Blazhko cycle. One of the most exciting new findings is the period doubling phenomenon \citep{kol10a}, never detected from ground-based data. It occurs at some phases of the Blazhko cycle and manifests in an alternating shape of the light variation in consecutive pulsation cycles \citep[see also][]{szabo10}. It was not detected in any non-modulated RR~Lyrae star so far. Satellite data allow to study the light curve changes over the Blazhko cycle in detail. Clearly, consecutive Blazhko cycles are not exactly repetitive \citep{kol11}. Also, the period of the Blazhko modulation differs from cycle to cycle. In the frequency spectra, the Blazhko phenomenon manifests itself as equidistantly spaced multiplets around main pulsation frequency and its harmonics. Triplets and quintuplets are detected in ground-based observations \citep[see e.g.,][]{jj08,kolea09}. In satellite data, not only triplets and quintuplets are visible, but also tenth-order side peaks \citep{cea10}.

All the newly discovered features of the Blazhko effect put stringent constraints on the models proposed to explain this longstanding enigma. In fact, the two most elaborated models, the Magnetic Oblique Rotator/Pulsator model \citep[MORP,][]{sh00}, and the Non-radial Resonant Rotator/Pulsator model \citep[NRRP, e.g.,][]{VHDK98,ND01,dm04} are ruled out by observations for several reasons. First, strong dipole magnetic fields, necessary for the MORP model to work, were not detected in RR~Lyrae stars \citep{cea04,kb09}. Secondly, both the MORP and NRRP models predict that the observed light curve modulation should manifest through specific features in the frequency spectra. The dipole geometry of the magnetic field, considered in the MORP model, leads to a quintuplet structure in the frequency spectra. A more complicated field geometry, that could possibly lead to higher order modulation components was not considered so far. Within the NRRP model it is shown that the modes of $l=1$ are most easily excited, giving rise to a triplet structure in the frequency spectra. The excitation of higher order nonradial modes is less likely. In addition, due to cancellation effects, the expected amplitudes would be very low for high-order modes. To the contrary, in satellite data we clearly detect higher-order modulation components.  \cite{cea10} detected modulation components up to the tenth order. In both the MORP and NRRP models, strong asymmetries in the amplitudes of the side peaks, as observed in the majority of  Blazhko stars, are not trivial to reproduce. Considering the triplets, in about 75 per cent of the Blazhko variables, the higher frequency side peak has a higher amplitude than the lower frequency side peak \citep{al03}. Third, both models propose mechanisms that imply a clockwork regularity, predicting robust modulation periods, e.g., equal to the rotation period of the star, and repeatable Blazhko cycles. Neither is seen in the data. The Blazhko variation can change considerably even from cycle to cycle \citep{kol11}. In several stars two modulation periods are apparent \citep[e.g.,][]{ds73,lc04,szf07}. Hence, the connection between the period of the Blazhko modulation and the rotation period is highly questionable. Also, in several cases consecutive Blazhko cycles differ significantly, which manifests, e.g., in systematic growth or decay of the modulation amplitude.

We also mention the recent studies of the Blazhko phenomenon by
\citet{bk11} and \citet{jj11}. Using the amplitude equation
formalism \citep[see e.g.,][]{bg84}, \cite{bk11} showed that the high
order half-integer resonance, proposed by \cite{szabo10} to explain
the period doubling, can also cause the phase and amplitude
modulation as observed in Blazhko variables. The result is very
exciting and a more detailed analysis and confirmation through the
hydrodynamic modelling would be of great value. \cite{jj11} studied
the Blazhko variables in the globular cluster M5. They showed that
the Blazhko stars tend to be situated on the zero-age horizontal
branch and at the blue edge of the fundamental mode instability
strip. They speculate that this location hints that the Blazhko
effect may be connected to the mode switch from the fundamental mode
to the first overtone pulsation during the evolution.

Very recently, an idea proposed by \citet{st06,st10}, which connects the Blazhko effect with variable magnetic stellar cycles, has gained popularity.  In this idea, the Blazhko modulation is connected to the cyclic strengthening and weakening of turbulent convection in the outer stellar layers, caused by the postulated transient magnetic field. The field decays cyclically and is subsequently built up anew by the turbulent/rotational dynamo. Details of the process, particularly the interaction between magnetic filed, turbulent convection and pulsation were not discussed. Also, numerical estimates, e.g., on the expected strength of modulation or the necessary strength of the variable magnetic field, were not presented. For a critical analysis of the Stothers idea we refer the reader to \citet{GezaSF}.

In this paper we use our pulsation hydrocodes to investigate whether variable turbulent convection may cause such a light variation as is associated with the Blazhko effect. The variation of turbulent convection is put into the hydrodynamical models in an {\it ad hoc} way -- interaction with the magnetic field is neglected. Currently, there are no models capable of reproducing a variable magnetic field due to dynamo mechanisms and of describing its dynamical coupling with turbulent convection and highly non-linear pulsation, features we deal with in the case of RR~Lyrae variables.  As a consequence, our model is strongly simplified and our results will serve for qualitative comparison with the observations. Having this in mind, we state the premise of our paper. If we can reproduce the most important observational constraints, the Stothers idea is certainly worth further, more detailed investigation. But if we do not succeed it needs revision.

The structure of the paper is the following. In
Section~\ref{sec.methods} we present the details of our numerical
model testing the idea proposed by Stothers (2006). In
Section~\ref{sec.results} we describe the properties of our models
and compare our results with observations, focusing on overall
properties of Blazhko variables and on recent {\it Kepler}
observations of the famous Blazhko variable, RR~Lyr (KIC 7198959)
\citep{kol11}. Some additional models are discussed in
Section~\ref{sec.discussion} and conclusions are presented in
Section~\ref{sec.conclusions}.

\section{Numerical methods}\label{sec.methods}

The basic tool in our modelling is the Warsaw non-linear convective pulsation hydrocode \citep{sm08a}, which we briefly describe in Section~\ref{sec.code}. The code is slightly modified in order to model the effects of variable turbulent convection. Details, as well as the modelling procedure, are outlined in Section~\ref{sec.idea}. In Section~\ref{sec.cms} we present the computed sequences of models to be analysed in this paper.

\subsection{Non-linear convective hydrocodes}\label{sec.code}

In all our computations we use pulsation codes described by \citet{sm08a}. These are a static model-builder, linear non-adiabatic code and a direct time integration non-linear hydrocode.  For convective energy transfer we use the \citet{ku86} convection model reformulated for the use in stellar pulsation codes. Radiation is described in the diffusion approximation. The codes use a simple Lagrangian mesh.

For an extensive description and details of numerical implementation we refer the reader to \citet{sm08a}. Here we note that the model equations contain eight order-of-unity scaling parameters that describe the turbulent convection model. These are mixing-length, $\alpha$, and parameters multiplying the turbulent fluxes and terms that drive/damp the turbulent energy, $\alpha_{\rm p}$ (turbulent pressure), $\alpha_{\rm m}$ (eddy-viscous dissipation), $\alpha_{\rm c}$ (convective heat flux), $\alpha_{\rm t}$ (kinetic turbulent energy flux), $\alpha_{\rm s}$ (buoyant driving), $\alpha_{\rm d}$ (turbulent dissipation) and $\gamma_{\rm r}$ (radiative cooling). Theory provides no guidance for their values. However, some standard values are in use, which result from comparison of a static, time-independent version of the model with the standard mixing-length theory \citep[see][]{wf98,sm08a}. In practice, values of these parameters should be such that models satisfy as many observational constraints as possible.

We also stress that we use the original \cite{ku86} prescription, in which essential buoyancy effects are included in convectively stable regions of the model (negative buoyancy). We note that the neglect of negative buoyancy \citep[as is done, e.g., in the Florida-Budapest code,][]{koea02} can lead to unphysical effects in the computed models. For an extensive discussion on the subject we refer the reader to, e.g., \citet{sm08b} or \citet{rs09}.

\subsection{Exploring Stothers' idea}\label{sec.idea}

\citet{st06} proposed that the Blazhko modulation is connected to the postulated cyclic strengthening and weakening of turbulent convection in the outer stellar layers, caused by a transient magnetic field. Turbulent convection becomes more vigorous during the decay of the magnetic field and it is quenched as the magnetic field builds up anew. Details of the underlying processes, particularly the interaction between magnetic field, turbulent convection and pulsation were not elaborated by Stothers.

\begin{figure}
\resizebox{\hsize}{!}{\includegraphics{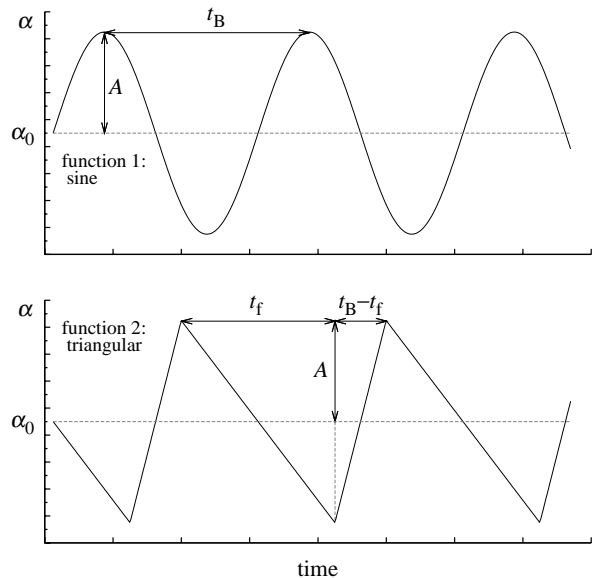}}
\caption{Possible shapes of mixing-length modulation considered in this study.}
\label{fig.functions}
\end{figure}

 Our code neglects the effects of magnetic fields and as such is not suitable for modelling the dynamical coupling of pulsation and turbulent convection on the one hand, and the transient magnetic fields postulated by \citet{st06} on the other hand. We only assume that the strength of the turbulent convection varies in time and do not elaborate on the physical mechanism behind such changes. The strength of turbulent convection is changed by cyclic variation of one of the $\alpha$-parameters entering our model (see Section~\ref{sec.code}). The mixing length parameter, $\alpha$, is our first choice, as it regulates the overall strength of convection and affects all the turbulent quantities entering the model \citep[see equations in e.g.,][]{sm08a}. We assume that the mixing length is either a sinusoidal or a triangular function of time, as illustrated in Fig.~\ref{fig.functions}.

The initial steps in our modelling correspond to a standard
hydrodynamic model computation, without modulation of the turbulent
convection (constant mixing-length). First, we construct a static
equilibrium model.  Next, we compute the non-linear model. The
static model is perturbed with a scaled linear velocity
eigenfunction followed by time evolution of the model. The model
integration is stopped, when the full-amplitude single-periodic
pulsation (the limit cycle) is reached. Then we start to modulate
the turbulent convection. The non-linear model integration is
restarted with the mixing-length parameter varying in time according
to the chosen functional form (see Fig.~\ref{fig.functions}). The
model is integrated for several Blazhko cycles. After a few initial
cycles, the consecutive ones are almost indistinguishable from one
another, indicating that the Blazhko limit cycle is reached.  Then,
the model integration is stopped and the resulting light variation
is subjected to a detailed analysis (Section~\ref{sec.results}).
\footnote{We note that our model predicts strictly periodic
Blazhko cycles, which result from our assumption of a strictly
periodic modulation of the mixing-length. This is not a necessary
feature of the model. We could easily modulate the mixing-length in
a quasi-periodic manner (as proposed in Stothers' paper), obtaining
a quasi-periodic modulation. We have chosen periodic modulation for
simplicity.}

In our computations we start to modulate the turbulent
convection at the phase of maximum radius -- the initial phase of
modulation relative to the unperturbed mono-mode pulsation is equal
to zero. To check whether the choice of the initial phase affects the
results, we have computed several additional models with different
initial phases: 0.25, 0.50 and 0.75, and starting the modulation of
turbulent convection with an either increasing or decreasing
mixing-length parameter. The final pulsation state (the Blazhko
limit cycle) is insensitive to the initial phase. The computed light
curves are almost indistinguishable for the models with different
initial phases (the trajectories overlap in the Fourier parameter plots,
Figs.~\ref{fig.papa}--\ref{fig.paps}). The amplitudes of the peaks in
the frequency spectra (see Section~\ref{sec.resfa}) are also
identical to within a fraction of a per cent.

\begin{figure}
\resizebox{\hsize}{!}{\includegraphics{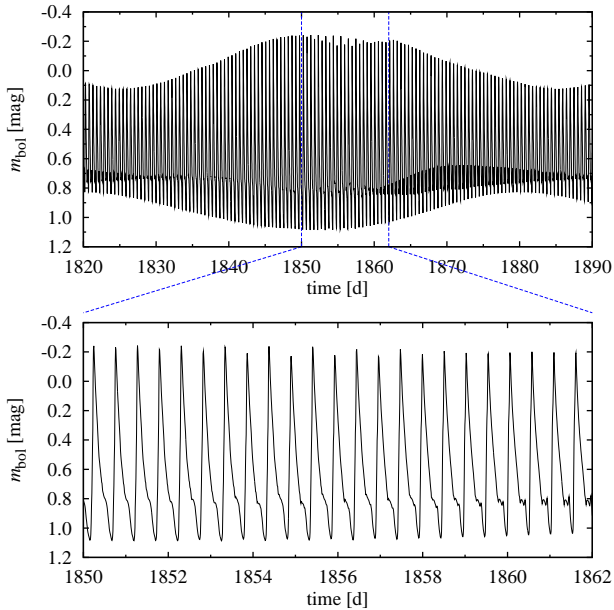}}
\caption{Bolometric magnitude versus time for a particular model of set M6 ($\teffs=6800\Ke$). In the lower panel we show the close-up of phases at which period doubling occurs.}
\label{fig.6800time}
\end{figure}

 In Fig.~\ref{fig.6800time} we present the light variation for a
typical model over slightly more than one Blazhko cycle. The
modulation of the pulsation amplitude is clearly visible. In the
lower panel we present the close-up of maximum amplitude phases at
which the period doubling occurs.\footnote{{\it See also the
animated gif available in the electronic version of the Journal. It
shows the light curve variation through the Blazhko cycle, including
the period doubling phenomenon, as well as the variation of the
Fourier parameters.}}

\subsection{Computed model sequences}\label{sec.cms}

In our modelling we focus on the fundamental mode models only, as most Blazhko variables are fundamental mode pulsators. 
The chemical composition of all our models is the same, $X=0.76$ and $Z=0.001$, which is typical for RR~Lyrae stars. We note that the metallicities of the Blazhko variables do not differ from those of the non-modulated RR~Lyrae pulsators \citep{rs05}. In all model computations we use OP opacities \citep{sea05} at low temperatures supplemented with the \citet{af94} opacity data. Opacities were computed for the solar mixture of \citet{a04}. Other physical parameters of the initial, non-modulated models (masses and luminosities) are collected in Table~\ref{tab.phys.nonmodulated}. Parameters of set A, which is the most explored set in this study, are close to those quoted for RR~Lyr \citep{kol10b}. In sets B, C and D the luminosity is varied, which allows to study the effects of a different $M/L$-ratio.

\begin{table}
\centering \caption{Physical parameters of the computed sequences of
initial models. The chemical composition of all our models is
the same, $X=0.76$ and $Z=0.001$. Effective temperatures are in the
range of $6300-7000$\thinspace K.}
\label{tab.phys.nonmodulated}
\centering
\begin{tabular}{ccccc}
\hline
Set  & $M\,[{\rm M}_\odot]$ & $L\,[{\rm L}_\odot]$\\
\hline
A   & 0.65 & 50.0 \\
B   & 0.65 & 40.0 \\
C   & 0.65 & 60.0 \\
D   & 0.65 & 70.0 \\
\hline
\end{tabular}
\end{table}

The convective parameters of the initial non-modulated models are
collected in Table~\ref{tab.convpar.nonmodulated}. We adopt only one
set of convective parameters, very similar to the one we have
adopted in \citet{bar09} and \citet{rs09}. With these convective
parameters, overall properties of the Galactic Cepheids, both
fundamental mode and first overtone pulsators, were modelled
successfully. Here, we only increased the eddy-viscous dissipation
($\alpha_{\rm m}$ parameter) in order to match the typical pulsation
amplitudes of the RR~Lyrae stars. We note that the modelling of
RR~Lyrae light curves is a difficult problem and some discrepancies
with the observations still remain (see e.g., the detailed comparison
done by \cite{kk98} or \cite{f99}). Nevertheless, the general
properties of the RR Lyrae light curves are quite well reproduced with the
current convective hydrocodes. In particular, the light curves of individual stars can be nicely matched with the hydrodynamic models \citep[see e.g.,][]{mar09,marc05}.

\begin{table}
\caption{Convective parameters considered for non-modulated initial models. Parameters $\alpha_{\rm s}$, $\alpha_{\rm c}$, $\alpha_{\rm d}$, $\alpha_{\rm p}$ and $\gamma_{\rm r}$ are given in the units of standard values \citep[$\alpha_{\rm s}=\alpha_{\rm c}=1/2\sqrt{2/3}$, $\alpha_{\rm d}=8/3\sqrt{2/3}$, $\alpha_{\rm p}=2/3$ and $\gamma_{\rm r}=2\sqrt{3}$; see][for details]{sm08a}.}
\label{tab.convpar.nonmodulated}
\centering
\begin{tabular}{cccccccc}
\hline
$\alpha$ & $\alpha_{\rm m}$ & $\alpha_{\rm s}$ & $\alpha_{\rm c}$ & $\alpha_{\rm d}$ & $\alpha_{\rm p}$ & $\alpha_{\rm t}$ & $\gamma_{\rm r}$ \\
\hline
1.5 & 0.65 & 1.0 & 1.0 & 1.0 & 0.0 & 0.0 & 1.0\\
\hline
\end{tabular}
\end{table}

The static models, the first step in our modelling, consist of 150 mass zones extending down to $2\cdot 10^6\,{\rm K}$. Fifty outer zones have equal mass, down to the anchor zone in which the temperature is set to $11000\,{\rm K}$ (hydrogen ionisation). The mass of the remaining zones increases geometrically inward. In the second step of our modelling procedure, we have computed non-linear full amplitude models. To reach the limit-cycle pulsation, we have computed 2000 pulsation cycles by default. These non-linear non-modulated models were subsequently used as initial models for model integrations with variable turbulent convection.

Several sequences of modulated models were computed. The properties of these models are summarised in Table~\ref{tab.modul}. We investigated the effects of different amplitudes of the turbulent convection modulation, different modulation periods, and different modulation shapes (see Fig.~\ref{fig.functions}). In each set, several models of different effective temperatures, extending through the whole instability strip, were computed. For most of the computed models, the physical parameters of set A were used (Table~\ref{tab.phys.nonmodulated}). We note that the mean physical parameters of the modulated models, such as mean radius or mean effective temperature, are affected by the modulation of turbulent convection and differ from the static equilibrium values. In the following, we will refer to the particular models by providing the adequate set name from Table~\ref{tab.modul} and the effective temperature of the underlying static model, $\teffs$.

\begin{table}
\caption{Parameters of turbulent convection modulation for the model sequences considered in this study. For definitions of $A$, $t_{\rm B}$ and $t_{\rm f}$ see Fig.~\ref{fig.functions}.}
\label{tab.modul}
\centering
\begin{tabular}{cccccc}
\hline
\multirow{2}{*}{Set} & physical   & \multirow{2}{*}{function} & \multirow{2}{*}{$A$} & \multirow{2}{*}{$t_{\rm B}$} & \multirow{2}{*}{$t_{\rm f}$}\\
                     & params. & & &\\
\hline
M1   & A & sine & 10\% & 60\thinspace d&-\\
M2   & A & sine & 20\% & 60\thinspace d&-\\
M3   & A & sine & 30\% & 60\thinspace d&-\\
M4   & A & sine & 20\% & 40\thinspace d&-\\
M5   & A & sine & 40\% & 60\thinspace d&-\\
M6   & A & sine & 50\% & 60\thinspace d&-\\
M7   & A & sine & 50\% & 40\thinspace d&-\\
M8   & A & sine & 50\% & 120\thinspace d&-\\
\hline
M7L4 & B & sine & 50\% & 40\thinspace d&-\\
M7L6 & C & sine & 50\% & 40\thinspace d&-\\
M7L7 & D & sine & 50\% & 40\thinspace d&-\\
\hline
MT1  & A & triang. & 50\% & 60\thinspace d & 45\thinspace d \\
MT2  & A & triang. & 50\% & 60\thinspace d & 15\thinspace d \\
\hline
\end{tabular}
\end{table}

\section{Results}\label{sec.results}

In this section we analyse the computed model sequences, described in Section~\ref{sec.cms}, focusing on the light curve variation through the Blazhko cycle, and on the properties of the frequency spectra. First, in Section~\ref{sec.reslcv}, we describe the overall properties of the light curve modulation in terms of the Fourier decomposition parameters \citep[e.g.,][]{sl81}. The variety of behaviours and trends we compute can serve for comparison with continuous observations of Blazhko variables by satellite missions. In Section~\ref{sec.resrr} we compare our results with {\it Kepler} observations of RR~Lyr, for which a detailed analysis of the light variation was already done \citep{kol11}. In Section~\ref{sec.ppc} we briefly analyse the pulsation period changes in our models and in Section~\ref{sec.resfa} we present the results of the frequency analysis of our models, focusing on the overall properties of the frequency spectra and how these compare with properties of Blazhko variables and RR~Lyr in particular. Finally, in Section~\ref{sec.PD} we provide a more detailed discussion on the period doubling phenomenon.

\subsection{Light curve variation through the Blazhko cycle}\label{sec.reslcv}

To study the light curve variation through the Blazhko cycle we use the time dependent Fourier analysis \citep{tdf}. We fit the Fourier series to the light variation of the consecutive pulsation cycles,
\begin{equation} m=A_0+\sum_{j}A_j\sin(j\omega t+\varphi_j),\end{equation}
and analyse time variation of the Fourier decomposition parameters of low order, the amplitude, $A_1$, amplitude ratio, $R_{21}$, and phase difference, $\varphi_{21}$,
\begin{equation}R_{21}=A_2/A_1,\ \ \varphi_{21}=\varphi_2-2\varphi_1.\end{equation}
We note that the derivation of instantaneous amplitudes and phases through the time-dependent Fourier analysis is justified, as the modulation is slow compared with the period of oscillation.

\begin{figure*}
\centering
\resizebox{\hsize}{!}{\includegraphics{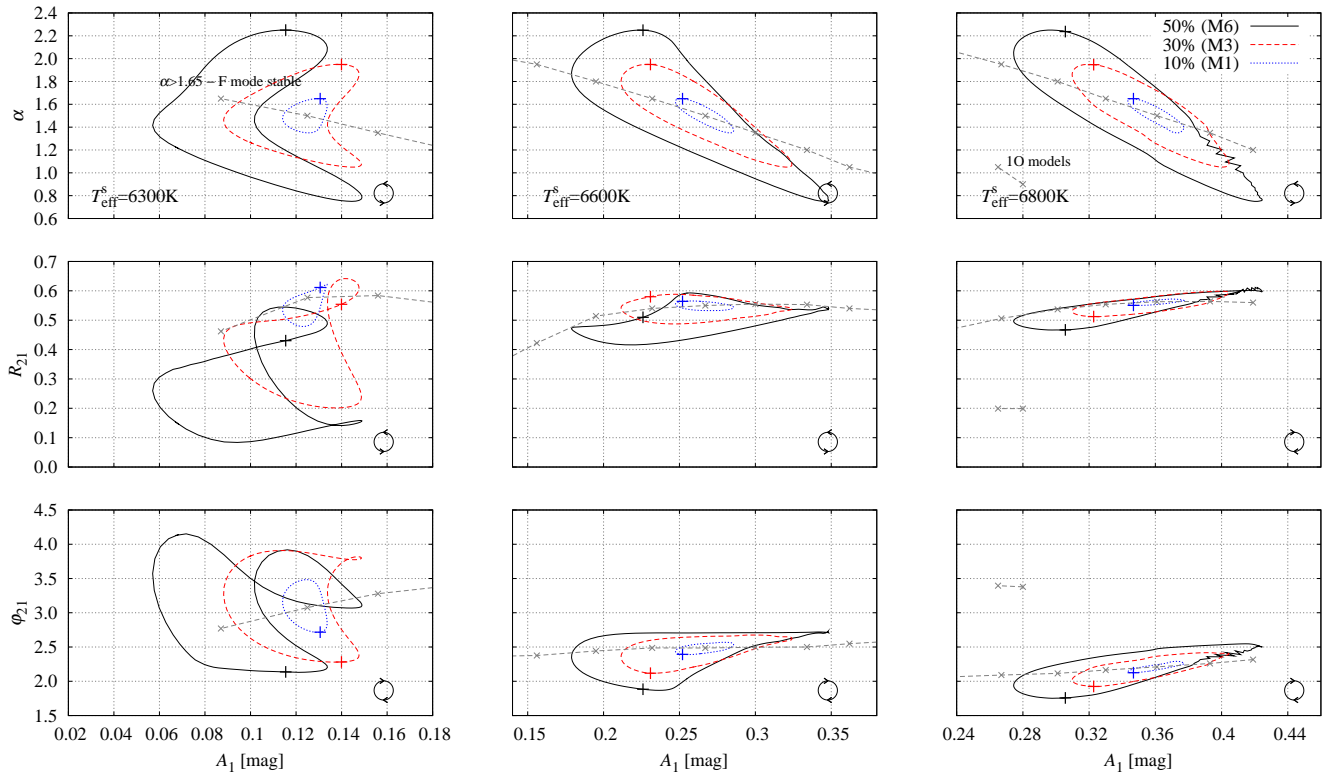}}
\caption{Light curve variation through the Blazhko cycle for the models with different amplitude of mixing-length modulation, 10 per cent (set M1), 30 per cent (set M3) and 50 per cent (set M6). The instantaneous mixing-length,  $R_{21}$ and $\varphi_{21}$ are plotted versus $A_1$ in the top, middle and bottom rows of the Figure. In consecutive columns we plot the results for models of different \teffs{} (6300\Ke, 6600\Ke{} and 6800\Ke). A plus sign on each trajectory indicates the phase of maximum mixing length. The direction of the time flow along each trajectory is shown on a schematic circle. Crosses connected with a long-dashed line correspond to the non-modulated models with different values of the mixing length. For \teffs=6300\Ke{} and $\alpha>1.65$ the fundamental mode is linearly stable and the models were not computed. For \teffs=6800\Ke{} and $\alpha\le 1.05$ the models switch into first overtone (1O) pulsation (two models in the Figure). }
\label{fig.papa}
\end{figure*}

\begin{figure*}
\centering
\resizebox{\hsize}{!}{\includegraphics{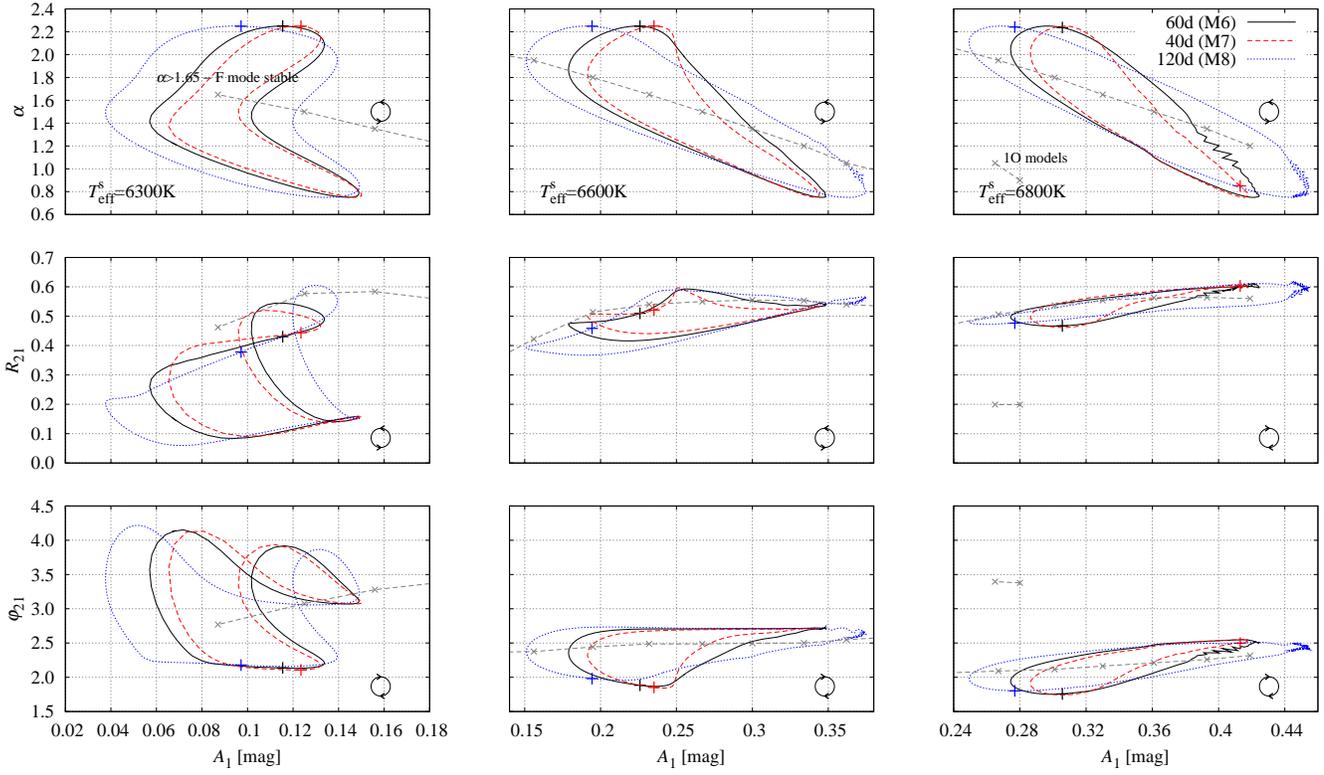}}
\caption{The same as Fig.~\ref{fig.papa}, but for the models with different periods of mixing-length modulation; 60\thinspace days (M6), 40\thinspace days (M7) and 120\thinspace days (M8).}
\label{fig.papp}
\end{figure*}

\begin{figure*}
\centering
\resizebox{\hsize}{!}{\includegraphics{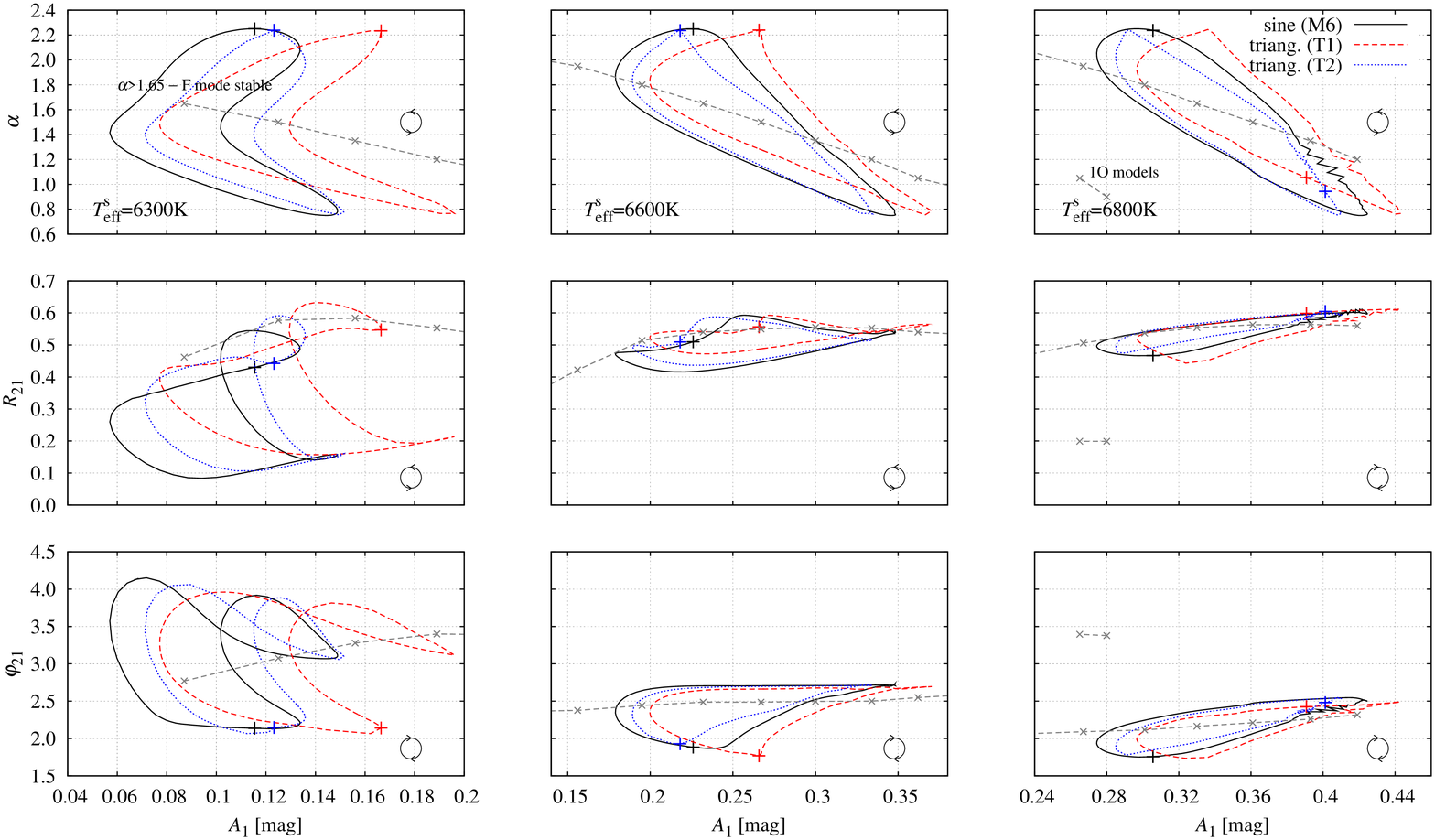}}
\caption{The same as Fig.~\ref{fig.papa}, but for the models with different shapes of mixing-length modulation; sinusoidal (set M6) and asymmetric triangular (MT1 and MT2).}
\label{fig.paps}
\end{figure*}

For the analysed smooth hydrodynamical data, the Fourier parameters also vary smoothly with time. If period doubling occurs, the alternating shapes of the consecutive pulsation cycles manifest in series of wiggles that appear in the run of the Fourier parameters. In Figs.~\ref{fig.papa}--\ref{fig.paps} we plot different relations between the Fourier parameters for different models. We present the effects of different amplitudes of the mixing-length modulation (Fig.~\ref{fig.papa}), different modulation periods (Fig.~\ref{fig.papp}) and different modulation shapes (Fig.~\ref{fig.paps}) on the light curve variation. The instantaneous mixing length, $\alpha$, $R_{21}$ and $\varphi_{21}$, are plotted versus the amplitude, $A_1$, in the top, middle and bottom rows of these figures, respectively. As in our models consecutive Blazhko cycles are repetitive, relations take the form of closed trajectories. In order to visualise the variation across the instability strip, we display the models with different effective temperatures, for which we choose \teffs=6300\Ke{} (left columns), \teffs=6600\Ke{} (middle columns; models to be compared later with RR~Lyr) and \teffs=6800\Ke{} (right columns; models with period doubling). All models are characterised by the physical parameters of set A (Table~\ref{tab.phys.nonmodulated}). On each trajectory in these figures, a plus sign indicates the phase of maximum mixing length, and the direction of the time flow along the trajectory is depicted schematically. In addition, in each panel crosses connected with dashed line indicate the location of several non-modulated models with fixed mixing-length values, in a range covered by our modulated models. Below we describe the properties of the light curve variation connected with the location within the instability strip and different parameters of the modulation of the turbulent convection.


\subsubsection{General properties of the light curve modulation. Variation through the instability strip.}
Analysis of Figs.~\ref{fig.papa}--\ref{fig.paps} reveals some general properties of the computed trajectories, independent of the parameters of turbulent convection modulation (amplitude, period and shape). A basic observation is that the higher the effective temperature is, \teffs, the higher is the mean pulsation amplitude, $A_1$. Also, the higher the mean pulsation amplitude, the smaller the range of variation of $R_{21}$ and $\varphi_{21}$. For the models with \teffs=6300\Ke{} we are close to the linear red edge of the fundamental mode. In fact, for the models with a strong modulation of the turbulent convection (e.g., of the sets M3 and M6 in Fig.~\ref{fig.papa}), phases of large instantaneous mixing-length values ($\alpha>1.65$) correspond to the pulsationally stable equilibrium models. On the other hand, at higher temperatures (\teffs=6800\Ke), small values of the mixing length ($\alpha\le 1.05$) at some modulation phases, if adopted in non-modulated models, would lead to first overtone (1O) pulsation (see the long-dashed lines in the figures). It is clear that the temporary light curve of the model in which turbulent convection is modulated is significantly different from that of the non-modulated model adopting the same value of the mixing length. In addition, hysteresis is clearly visible for the modulated models. Very different shapes of the light curve are possible at two phases characterised by the same instantaneous mixing length. The described features are in qualitative agreement with the observations. \cite{jbs02} noted that the light curves of Blazhko stars are never (at any phase of the Blazhko cycle) like those of non-Blazhko stars, thus always distorted. A detailed analysis of the light curve variation in {\it Kepler} RR~Lyr data \citep{kol11} reveals a similar behaviour of the Fourier parameters as plotted in the figures. A more quantitative comparison is postponed to the next section.

Another interesting feature is visible in the $\alpha$ vs. $A_1$
plots (top rows of Figs.~\ref{fig.papa}--\ref{fig.paps}). We note
that the minimum value of the amplitude does not coincide with the maximum
value of the mixing-length parameter, as one might naively expect,
but it is delayed. The reason for such a delay is not clear. It
shows the inertia of the dynamical system, which does not adjust
instantaneously to the changing dissipation.

Considering the relations $R_{21}$ vs. $A_1$ and $\varphi_{21}$ vs. $A_1$ (middle and bottom rows of Figs.~\ref{fig.papa}--\ref{fig.paps}) we see that for low temperature models (of small mean pulsation amplitudes) trajectories have the shape of double loops, while at higher temperatures (and high mean pulsation amplitudes) single loops are present. The direction of trajectories (in case of double loops, the direction of the larger loop) is always clockwise for $\varphi_{21}$ vs. $A_1$ trajectory (bottom rows of Figs.~\ref{fig.papa}--\ref{fig.paps}). For $R_{21}$ vs. $A_1$ trajectories (middle rows of Figs.~\ref{fig.papa}--\ref{fig.paps}) the direction is either counter-clockwise (for cooler models, \teffs=6300\Ke{} and \teffs=6600\Ke) or clockwise (for the hottest models, \teffs=6800\Ke). The analysis of additional models across the instability strip reveals that the transition occurs at around \teffs=6700\Ke{}. For these models we deal with an almost single-valued dependence of $R_{21}$ on $A_1$. We note that the direction of the $\varphi_{21}$ vs. $A_1$ loop is related to a particular asymmetry of the triplet components in the frequency spectra \citep{sj09}. We discuss this problem in more detail in Section~\ref{sec.fams}.

Considering the $R_{21}$ vs. $A_1$ relation, we first note that intuitively we expect that $R_{21}$ should correlate with $A_1$.  The higher the amplitude, $A_1$, the larger the non-linearity and consequently larger the contribution of the harmonic terms (and thus, higher $R_{21}$).  In our models, in which both $R_{21}$ and $A_1$ are functions of time, the relation between $R_{21}$ and $A_1$ is complicated. For the hottest models, of higher mean pulsation amplitude, $R_{21}$ correlates with $A_1$. For the lower temperature models, of lower mean pulsation amplitudes, the trajectories are more extended ("blown-up") and the relation is not obvious. At some phases in the Blazhko cycle, a high amplitude $A_1$ is accompanied by lower values of $R_{21}$.

Period doubling is clearly visible for the highest temperature models ($\teffs=6800$\Ke) displayed in Figs.~\ref{fig.papa}--\ref{fig.paps}, in which the mixing-length modulation is strong (Fig.~\ref{fig.papa}). During the phases with period doubling, the trajectories are no longer smooth. The series of wiggles is present, which indicates that the consecutive pulsation cycles alternate. Period doubling will be discussed in more detail in Section~\ref{sec.PD}.

In addition to the models running horizontally in the instability strip, we have computed some models with different $M/L$ ratio. Four sequences of models were computed, M7L4, M7, M7L6 and M7L7 (see Table~\ref{tab.modul}), in which the luminosities vary from $40\LS$ to $70\LS$. The mass is the same in all these model sequences ($0.65\MS$) and \teffs{} was varied. In all these sequences the amplitude of the mixing-length modulation was 50 per cent, the modulation period 40\thinspace days, and the shape of the modulation was sinusoidal. Except for the shifts in the computed trajectories, connected with the different luminosities of the initial models, the trajectories are very similar and hence not presented in a separate figure. We have noted that the direction of the $R_{21}$ vs. $A_1$ trajectories for high temperature models depends on the luminosity and is clockwise for the low-luminosity models ($40\LS$ and $50\LS$) and counter-clockwise for the high-luminosity models ($60\LS$ and $70\LS$), in which $R_{21}$ displays a flat dependence on $A_1$. Consequently, a clockwise direction of the $R_{21}$ vs. $A_1$ trajectory is present only in higher-temperature and lower-luminosity models.

\subsubsection{Different strength of the turbulent convection modulation.}
In Fig.~\ref{fig.papa} we plot the results for the models with a different amplitude of the turbulent convection modulation, $A=10$ per cent (M1), $A=30$ per cent (M3) and $A=50$ per cent (M6). The period and shape of the modulation are the same for these three sets ($P_{\rm B}=60\,{\rm d}$, sine). As expected, the stronger the modulation, the more complicated (`non-linear') trajectories we get and higher the ranges of variation of the Fourier parameters. This property of the computed trajectories will be used in the next section to estimate the required amplitude of the mixing-length modulation to reproduce the RR~Lyr observations. We also note that, depending on the strength of the modulation, the mean values of Fourier parameters vary. This concerns the amplitude, $A_1$, and the amplitude ratio, $R_{21}$. The mean phase difference is not very sensitive to the strength of the modulation. For the most convective models of \teffs=6300\Ke, the stronger the modulation, the smaller the amplitude and the amplitude ratio.

\subsubsection{Different period of the turbulent convection modulation.}
In Fig.~\ref{fig.papp} we plot the results for models with different period of turbulent convection modulation, $P_{\rm B}=40\,{\rm d}$ (M7), $P_{\rm B}=60\,{\rm d}$ (M6) and $P_{\rm B}=120\,{\rm d}$ (M8). The amplitude and shape of the modulation are the same for these three sets ($A=50$ per cent, sine). The basic observation is that for longer modulation period, the computed loops are larger (more "blown-up"), and thus, the larger the ranges of variation of Fourier parameters, but the effect is not as strong as in case of different amplitudes of turbulent convection modulation. We also note that the range of effective temperatures in which the period doubling occurs, depends on the modulation period, but we postpone the detailed discussion to Section~\ref{sec.PD}.

\subsubsection{Different shape of the turbulent convection modulation.}
In Fig.~\ref{fig.paps} we plot the results for models of sets M6 (sinusoidal modulation of the mixing length), MT1 and MT2 (asymmetric triangular modulation). The period and amplitude of mixing-length modulation are the same for these three sets ($P_{\rm B}=60\,{\rm d}$, $A=50$ per cent). We conclude that the differences between the trajectories are rather minute and the light curve variation is very similar for the different shapes of modulation.

\subsection{Comparison with RR~Lyr}\label{sec.resrr}

RR~Lyr is not only the prototype of the RR~Lyrae variables but also
a famous Blazhko variable with a strongly modulated light curve. Its
Blazhko period is subject to small variations, its present value
being $39.1\pm 0.3$ \thinspace days \citep{kol11}. Thanks to its
location in the {\it Kepler} field of view, we now have excellent,
nearly continuous photometric data covering slightly more than three
Blazhko periods -- seasons Q1+Q2 of the {\it Kepler} long cadence
data \citep{jen10a,jen10b} of which Q1 is already
public\footnote{\textsf{http://archive.stsci.edu/kepler/}}.

In this section we compare the light curve variation through the
Blazhko cycle, as it is observed in RR~Lyr, with our models.
The {\it Kepler} photometric passband is rather wide, covering the
combined range of the standard {\it V} and {\it R} passbands
\citep{koch}.
It justifies the direct comparison of our model bolometric light
curves with the {\it Kepler} RR~Lyr light curve. This simplification
does not affect the estimates presented in this section. We note
that \cite{jim} derived the transformation between the Fourier
parameters in {\it Kp} and in {\it V} passbands based on the
photometry of three non-modulated RR~Lyrae stars in both bands. The
differences are systematic but very small.

\begin{figure}
\resizebox{\hsize}{!}{\includegraphics{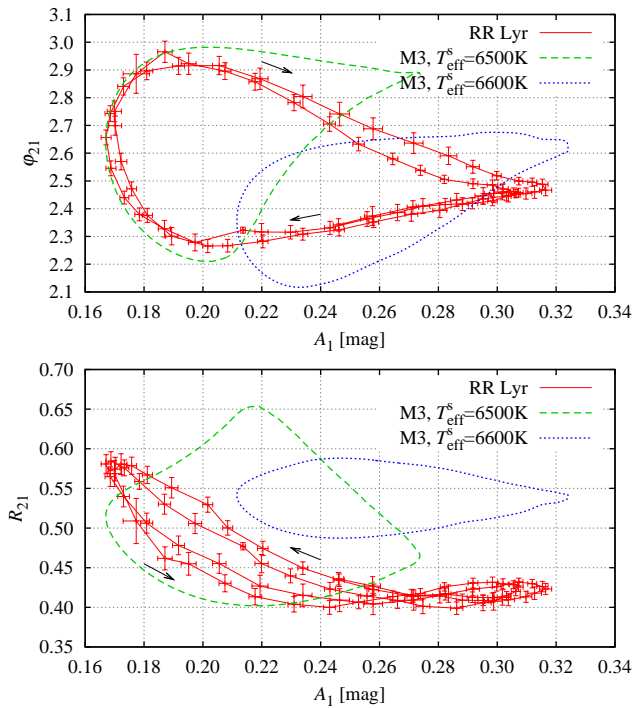}}
\caption{Fourier parameters, $\varphi_{21}$ and $R_{21}$, plotted vs. $A_1$ for {\it Kepler} RR~Lyr data (season Q2) and for two models of set M3.}
\label{fig.comRR}
\end{figure}

In Fig.~\ref{fig.comRR} we plot the $\varphi_{21}$ vs. $A_1$ and $R_{21}$ vs. $A_1$ relations as observed for RR~Lyr (season Q2 of {\it Kepler} data). The ranges of variation of the Fourier parameters are rather large. In our models, as noted in the previous section, the larger the amplitude of turbulent convection modulation is, the larger is the range of variation of the Fourier parameters. Now, we can estimate the strength of mixing-length modulation necessary to explain the RR~Lyr observations, through comparing the model and observed ranges of variation of $A_1$, $R_{21}$ and $\varphi_{21}$. To this aim, for each Fourier parameter $c$, $c\in\{A_1,\, R_{21},\, \varphi_{21}\}$, we define the parameter $\Delta c$,
\begin{equation}\Delta c=2\frac{c_{\rm max}-c_{\rm min}}{c_{\rm max}+c_{\rm min}}\,,\label{DeltaFourier}\end{equation}
constructed using the minimum, $c_{\rm min}$, and maximum, $c_{\rm max}$, values of $c$ during the modulation cycle. We note that the value of $\Delta c$ is independent of simple scaling of the {\it Kepler} data by a constant factor. For RR~Lyr we have $\Delta A_1=0.62$, $\Delta R_{21}=0.38$ and $\Delta\varphi_{21}=0.27$ \citep[][Fig.~\ref{fig.comRR}]{kol11}. The computed values for the models with a different strength of the turbulent convection modulation, $A=10$ per cent (set M1), $A=30$ per cent (set M3), and $A=50$ per cent (sets M6 and M7) are collected in Table~\ref{tab.RRcomparison}. For each set, the values for seven models of different \teffs{} are computed. The values that agree within 20 per cent with RR~Lyr values are marked with an asterisk. Note that the sets M6 and M7 have the same amplitude of the mixing-length modulation ($A=50$ per cent), but for set M7 the modulation period is shorter (40\thinspace days, very close to RR~Lyr's modulation period).

\begin{table*}
  \caption{Ranges of variation of the Fourier parameters for the different models of sets M1, M3, M6 and M7 compared with the RR~Lyr {\it Kepler} data. In the first column effective temperature of the initial non-modulated model, \teffs, is given. In the consecutive columns the ranges of variation of Fourier parameters, $\Delta A_1$, $\Delta R_{21}$ and $\Delta\varphi_{21}$ (see eq.~\ref{DeltaFourier}) are given for different sets of modulated models. In the last row, the data for RR~Lyr are provided for reference. Model values that agree with the corresponding RR~Lyr value within 20 per cent are marked with an asterisk.}
\label{tab.RRcomparison}
\centering
\begin{tabular}{l||rrrc|rrrc|rrrc|rrr}
& \multicolumn{4}{|c}{M1 ({\it A}=10\%)}&\multicolumn{4}{|c}{M3 ({\it A}=30\%)}&\multicolumn{4}{|c}{M6 ({\it A}=50\%)}&\multicolumn{3}{|c}{M7 ({\it A}=50\%)}\\
\hline
 $\teffs$& $\Delta A_1$ & $\Delta R_{21}$ & $\Delta \varphi_{21}$&& $\Delta A_1$ & $\Delta R_{21}$ & $\Delta \varphi_{21}$&& $\Delta A_1$ & $\Delta R_{21}$ & $\Delta \varphi_{21}$&& $\Delta A_1$ & $\Delta R_{21}$ & $\Delta \varphi_{21}$\\
\hline
6300K & 0.16 & 0.26 & *0.25 &&  *0.51 & 1.05 &  0.53 &&  0.89 &  1.47 & 0.64 &&      0.79 &  1.41 & 0.65 \\
6400K & 0.16 & 0.27 &  0.17 &&  *0.53 & 0.78 &  0.39 &&  0.84 &  1.08 & 0.51 &&     *0.74 &  0.99 & 0.50 \\
6500K & 0.17 & 0.19 &  0.10 &&   0.48 & 0.48 & *0.30 &&  0.76 &  0.73 & 0.41 &&     *0.66 &  0.65 & 0.41 \\
6600K & 0.14 & 0.06 &  0.07 &&   0.42 & 0.19 & *0.23 && *0.64 & *0.35 & 0.37 &&     *0.57 & *0.30 & 0.38 \\
6700K & 0.11 & 0.02 &  0.07 &&   0.33 & 0.08 & *0.22 && *0.53 &  0.19 & 0.39 &&      0.46 &  0.17 & 0.40 \\
6800K & 0.09 & 0.04 &  0.08 &&   0.27 & 0.16 & *0.23 &&  0.43 &  0.28 & 0.37 &&      0.37 &  0.27 & 0.38 \\
6900K & 0.07 & 0.06 &  0.07 &&   0.21 & 0.17 & *0.22 &&  0.34 &  0.28 & 0.36 &&      0.32 &  0.28 & 0.36 \\
\hline
RR~Lyr &0.62 & 0.38 &  0.27 &&    0.62 & 0.38 &  0.27 &&   0.62 & 0.38 &  0.27 &&   0.62 & 0.38 &  0.27 \\
\hline
\end{tabular}
\end{table*}

To reproduce the ranges of the Fourier parameter variation in RR~Lyr, a mixing-length modulation with an amplitude equal to at least 30 per cent is necessary. The best match is found for an amplitude of the mixing-length modulation equal to 50 per cent (sets M6 and M7). Then, both $\Delta A_1$ and $\Delta R_{21}$ can be matched for the model with \teffs=6600\Ke. For models with \teffs=6600\Ke, the mean pulsation period ($\approx 0.573$\thinspace d) matches RR~Lyr's period \citep[0.567\thinspace d,][]{kol11} almost exactly. The range of the phase variation, $\Delta\varphi_{21}$, however, is slightly higher for these models than is observed in RR~Lyr.

The most stringent constraints on the Stothers model come from
highly modulated stars, like RR~Lyr. Tiny modulations as observed
for many Blazhko stars can be easily reproduced assuming a small
amplitude of turbulent convection modulation in our models. It is
now evident that, if the mechanism proposed by Stothers is
responsible for the Blazhko effect, the observed large modulations
of the light curves (as we observe in RR~Lyr) require significant
changes of the mixing-length over the Blazhko cycle.  Note that the
modulation amplitude, $A$, as defined in Fig.~\ref{fig.functions},
is actually a semi-amplitude. For the considered models, $A=50$ per
cent means that the mixing length, $\alpha$, varies in a huge range,
from 0.75 to 2.25. In our opinion, such large changes in the
effectiveness of the turbulent convection on relatively short
time-scales of typically tens to hundreds of days (typical Blazhko
periods) are highly unlikely (see discussion in
Section~\ref{sec.conclusions}).

Now we focus on the shape of the Fourier parameter trajectories. A close inspection of Figs.~\ref{fig.papa}--\ref{fig.paps}, particularly the models with \teffs=6600\Ke{} (for which the period of the fundamental mode matches that of RR~Lyr) shows that the model $\varphi_{21}$ vs. $A_1$ loops are quite similar to the RR~Lyr loop (top panel of Fig.~\ref{fig.comRR}). What we typically see in our models (around \teffs=6600\Ke) is a larger ("blown-up") part of the loops on the left side and cusps on the right side. This is what we observe in RR~Lyr, however, the cusp in our models is located at higher values of $\varphi_{21}$ compared to RR~Lyr. A comparison of {\it Kepler} RR~Lyr data with two models of set M3 (for which we get the best $\Delta\varphi_{21}$ match) is shown in the upper panel of Fig.~\ref{fig.comRR}. We note that also the direction of the time flow is the same for both RR~Lyr and for the models (clockwise).

Considering the $R_{21}$ vs. $A_1$ loops (bottom panel of Fig.~\ref{fig.comRR}) we cannot reproduce the behaviour that we observe in RR~Lyr. In RR~Lyr $R_{21}$ anticorrelates with $A_1$ through the majority of the Blazhko cycle. This is not reproduced by our models, although at some phases of the modulation $R_{21}$ increases as $A_1$ decreases. The bottom panel of Fig.~\ref{fig.comRR} provides a direct comparison of the model (set M3) and RR~Lyr loops. Also here, the direction of the time flow is the same for our models and for RR~Lyr (bigger loop, counter-clockwise).

\subsection{Changes of the pulsation period}\label{sec.ppc}

In this section we analyse the changes of the pulsation period.
In the mathematical description, a period change is equivalent to a
change of the pulsation phase. From the observational point of view, the
two are indistinguishable. It is the physical interpretation where
the difference occurs. Period changes are caused by the overall
changes of the stellar structure, while phase changes may result,
e.g., from the nonlinear interaction of pulsation modes (which does
not affect the stellar structure).

In our models, due to the changes of the convective structure, the
pulsation period changes during the modulation. The amplitude of
the period variation, $\delta P/P$, is most sensitive to the
amplitude of turbulent convection modulation, and hence can be used
to estimate the required strength of mixing-length modulation in a
similar way we have done in the previous section. In
Table~\ref{tab.ppc} we list the period changes for the models with
different amplitudes of the turbulent convection modulation, $A=10$
per cent (set M1), $A=30$ per cent (set M3) and $A=50$ per cent (set
M6). The period changes were estimated based on the radius variation
of our hydrodynamic models, using the time difference between
consecutive radius maxima as an estimate of the instantaneous
period. Then, the full amplitude of the period variation was used to
derive $\delta P/P$ values. It is evident that the larger the
amplitude of mixing-length modulation is, the larger the amplitude
of period variation. Also, the trend of decreasing amplitude of the
period variation with increasing effective temperature is clear. In
Blazhko RR~Lyrae variables, the pulsation period changes are clearly
detected, with a typical amplitude, $\delta P/P$, between 0.2 and
1.4 per cent \citep{mok10}. For RR~Lyr, the period change is 0.83
per cent \citep{kol11}. Comparison with values collected in
Table~\ref{tab.ppc} indicates that in order to reproduce such period
variations, large amplitudes of the mixing-length modulation are
necessary, at least of the order of 50 per cent, in agreement with
the value derived in the previous section and in agreement with
estimate of \cite{mok10}, based on the analysis of the periods of
linear equilibrium models.

\begin{table}
\caption{Amplitudes of the period change, $\delta P/P$, for several models of different \teffs{} and different amplitudes of turbulent convection modulation, 10 per cent (M1), 30 per cent (set M3) and 50 per cent (set M6).}
\label{tab.ppc}
\centering
\begin{tabular}{l|rrr}
 & \multicolumn{3}{|c}{$\delta P/P$ [per cent]}\\
\hline
 $\teffs$& M1 & M3 & M6\\
\hline
6300K & 0.12  & 0.39 & 0.97\\
6400K & 0.11  & 0.34 & 0.88\\
6500K & 0.092 & 0.31 & 0.77\\
6600K & 0.095 & 0.30 & 0.64\\
6700K & 0.051 & 0.28 & 0.51\\
6800K & 0.11  & 0.27 & 0.44\\
6900K & 0.10  & 0.20 & 0.49\\
\hline
\end{tabular}
\end{table}

\subsection{Analysis of frequency spectra}\label{sec.resfa}

In this section we focus on the frequency analysis of our models,
for which we use the \textsc{Period04} package \citep{period04}.
Differently from Section~\ref{sec.reslcv}, we now analyse the
light variation over many pulsation periods and many Blazhko cycles.
At this long time-scale the system is stationary and we can fit the
data with a Fourier sum of the following form:

\begin{eqnarray}
m&=&A_0+\sum_{k=1}^{N}\bigg\{A_k\sin\big(2\pi k f_0 t+\phi_k \big)+\nonumber \\
&&A_k^+\sin\Big[2\pi\big(kf_0+f_{\rm B}\big)t+\phi_k^+\Big]+\nonumber\\
&&A_k^-\sin\Big[2\pi\big(kf_0-f_{\rm B}\big)t+\phi_k^-\Big]+\nonumber\\
&&\sum_{i=2}^{J}A_k^{i+}\sin\Big[2\pi\big(kf_0+if_{\rm B}\big)t+\phi_k^{i+}\Big]+\nonumber\\
&&\sum_{i=2}^{J}A_k^{i-}\sin\Big[2\pi\big(kf_0-if_{\rm B}\big)t+\phi_k^{i-}\Big]\bigg\}+\nonumber\\
&&\sum_{j=1}^{L}B_j\sin\big(2\pi j f_{\rm B} t+\phi_{{\rm B},j} \big)\,.\label{eq.Fsum}
\end{eqnarray}
In the above formula, the first line corresponds to the main pulsation frequency, $f_0$, and its harmonics, $kf_0$, the second and third lines correspond to the components of the triplets, $kf_0\pm f_{\rm B}$, the next two lines describe other higher-order multiplet components and in the last line we account for the modulation frequency, $f_{\rm B}$, and its harmonics, $jf_{\rm B}$.  We assume that the components of the multiplets are equidistant. Also the modulation frequency, $f_{\rm B}$, is known, as it is equal to the inverse of the assumed period of mixing-length modulation (see Table~\ref{tab.modul}). Consequently, only one frequency is determined from the data, the main pulsation frequency, which is not known a priori (the non-linear period differs from the linear one). By default we use $N=19$, $J=5$ and $L=2$ for all the models (note that signal at $2f_{\rm B}$ is strong in our models).

The length of the hydrodynamic data we analyse corresponds to four full Blazhko cycles. To speed up the computations, the sampling of the model light curve is degraded to roughly 60 points per pulsation period ($\sim$twice the {\it Kepler} resolution for RR~Lyr).

Below, we present some details of our analysis for one particular model, and later (Section~\ref{sec.fams}) we discuss the general properties of the frequency spectra of our models and compare them with the observations.

\subsubsection{Frequency spectra analysis -- particular case}\label{sec.fapc}

We analyse one particular model from set M6 with $\teffs=6800\Ke$. The model displays a clear period doubling in the computed light curve (see, e.g., Fig.~\ref{fig.papa}). The mean pulsation period for this model is $P_1=0.517\,{\rm d}$, which is not far from the RR~Lyr period \citep[$P_1=0.567$\thinspace d,][]{kol11}. As noted before, the best match with RR~Lyr's period is obtained for the model with \teffs=6600\Ke{}, but this model does not display the period doubling.

For the discussed model, the variation of the bolometric magnitude with time for slightly more than one Blazhko cycle (70 days) is shown in the upper panel of Fig.~\ref{fig.6800time}. Period doubling is clearly visible during the phases of maximum pulsation amplitude. The lower panel of Fig.~\ref{fig.6800time} shows these phases. The period doubling is obvious and lasts for several pulsation cycles. However, its amplitude is not very large.

\begin{figure}
\resizebox{\hsize}{!}{\includegraphics{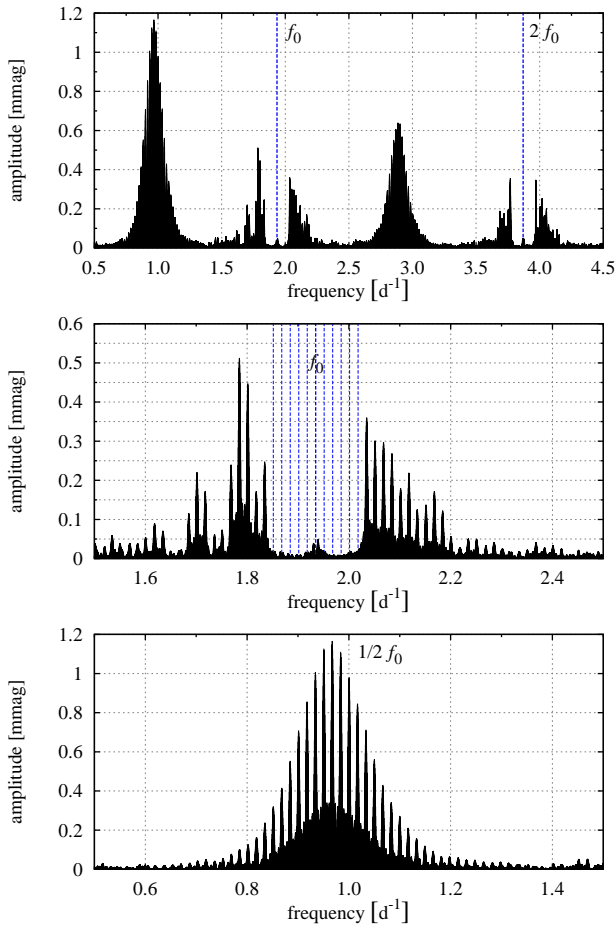}}
\caption{Frequency spectrum for the particular model of set M6 ($\teffs=6800\Ke$) after prewhitening with the Fourier sum described by Eq.~(\ref{eq.Fsum}) ({\it upper panel}). Two close-ups are shown: at the location of fundamental mode frequency ({\it middle panel}) and at its subharmonic, $1/2\,f_0$ ({\it bottom panel}). Removed frequencies are marked with dashed lines.}
\label{fig.6800spec}
\end{figure}

In Fig.~\ref{fig.6800spec} we plot some results of the frequency
analysis. The upper panel of Fig.~\ref{fig.6800spec} shows the
close-up around the fundamental mode frequency, $f_0$, and its
harmonic, $2f_0$, \mbox{after} prewhitening the data with the
fundamental mode frequency and its harmonics (dashed lines), and
five consecutive components of the multiplet structure. In the
middle panel of Fig.~\ref{fig.6800spec}, we show the vicinity of the
fundamental mode frequency. All removed frequencies are marked with
dashed lines.  Clearly, many more higher-order side peaks are
visible. Residual power at the position of $f_0$ corresponds to the
long-term evolution of the model toward the final limiting cycle
pulsation (which is of exponential character). The bottom panel of
Fig.~\ref{fig.6800spec} shows the vicinity of the half-integer
frequency (HIF), $1/2\,f_0$. The signal at this frequency is a
signature of the period doubling \citep[see][]{szabo10}, which
appears in the discussed model during a short fraction of the
modulation cycle. Consequently, the signal is strongly modulated
with the Blazhko period and hence many equally-spaced peaks are
clearly visible, with the separation corresponding to the modulation
frequency. The envelope of the signal located at the HIF's is very
regular and resembles a Gaussian. The width of this Gaussian
corresponds to the duration of the period doubling behaviour
observed in the model ($6-7$\thinspace days).

\begin{figure}
\resizebox{\hsize}{!}{\includegraphics{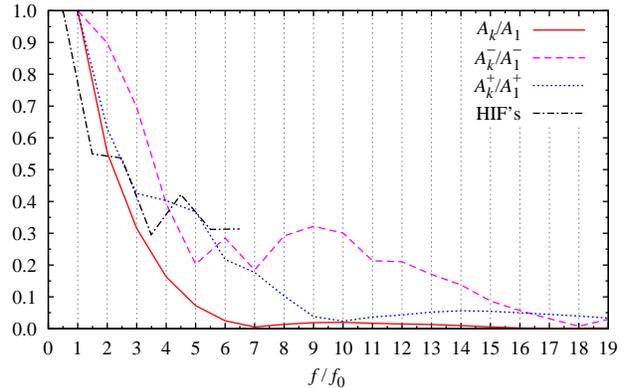}}
\caption{Amplitude ratios, $A_k/A_1$, $A_k^-/A_1^-$, $A_k^+/A_1^+$ and amplitude of the half-integer frequencies, plotted versus the frequency (harmonic orders indicated by vertical dashed lines). Results for the model of set M6 with $\teffs=6800\Ke$.}
\label{fig.6800ampy}
\end{figure}

Our analysis is focused on the first-order side peaks, the triplets. The relations between their amplitudes, $A_k^-$, $A_k$ and  $A_k^+$ (see Eq.~\ref{eq.Fsum}) were studied for several Blazhko stars observed from the ground \citep[e.g.,][]{jj06} as well as for the {\it Kepler} RR Lyr observations \citep{kol11}. For our models, the relations between the amplitudes of the signals at different frequencies are shown in Fig.~\ref{fig.6800ampy}. This is the analog of fig.~7 from \citet{kol11}, where the results for RR~Lyr are presented. The model we discuss now has physical parameters and a mean period close to RR~Lyr, but the modulation period is slightly longer (60\thinspace days). However, the described features of the frequency spectrum do not depend strongly on the modulation period and the analysis below is comparative.

For $A_k/A_1$ we observe an exponential decrease with harmonic order $k$. At $k=7$ the amplitude ratio drops to very small values (just like in RR~Lyr). For the modulation components the decrease of amplitude ratios ($A_k^-/A_1^-$ and $A_k^+/A_1^+$) is less steep. It is also less steep for  $A_k^-/A_1^-$ than for $A_k^+/A_1^+$ -- all in agreement with RR~Lyr. For $A_k^-/A_1^-$ we observe a tail at high $k$ also in agreement with RR~Lyr. For low $k$, $A_k^-/A_1^-$ increases (above 1) in the case of RR~Lyr. Here it is not the case, but such behaviour can be obtained for other models (see next section).

For the amplitudes of the HIF's resulting from period doubling, in the discussed model the highest peak is located at $1/2\,f_0$ followed by $3/2\,f_0$ and $5/2\,f_0$. For RR~Lyr the highest peak is observed at $3/2\,f_0$ followed by $5/2\,f_0$ and $1/2\,f_0$. We discuss this discrepancy in more detail in Section~\ref{sec.PD}.

\subsubsection{Frequency spectra analysis -- model sequences}\label{sec.fams}

\begin{figure}
\resizebox{\hsize}{!}{\includegraphics{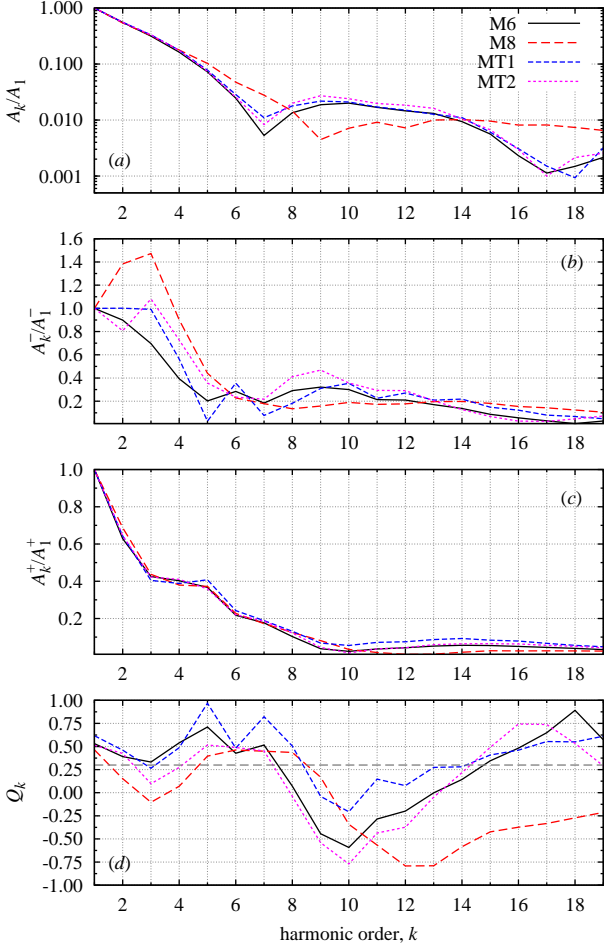}}
\caption{Relations between the amplitudes of the triplet components for the models of sets M6, M8, MT1 and MT2 (all with \teffs=6800\Ke). Panel ({\it a}): $A_k/A_1$ vs. $k$ in logarithmic scale; panel ({\it b}): $A_k^-/A_1^-$ vs. $k$; panel ({\it c}): $A_k^+/A_1^+$ vs. $k$, and panel ({\it d}): $Q_k$ vs. $k$. The distribution of $Q$ values from the MACHO database peaks at 0.3, which is marked with the horizontal dashed line for reference.}
\label{fig.ampFUN}
\end{figure}

A detailed frequency analysis was conducted for all the model sequences discussed in this paper. Typical results for models with \teffs=6800\Ke{} and different parameters of turbulent convection modulation are shown in  Fig.~\ref{fig.ampFUN}. These models are similar to that of set M6 discussed in the previous section, have the same amplitude of mixing-length modulation (50 per cent), but a longer modulation period (set M8) or a different shape of the modulation (triangular for MT1 and MT2). We stress that the results discussed below are characteristic for all our model sequences.

For the amplitudes of the harmonic frequencies, $A_k/A_1$ (panel {\it a} of Fig.~\ref{fig.ampFUN}), an interesting behaviour is observed at harmonic order around 7. The amplitudes are already very small at this order so the plots are in logarithmic scale to reveal the details. Local minima are clearly visible. They fall at $k=7$, except for the model with a longer modulation period (set M8, $k=9$). In all these models period doubling is present, however the origin of the discussed minima and their possible connection with period doubling is not clear (see also Section~\ref{sec.PD}).

Now we discuss the amplitudes of the modulation components, $A_k^-/A_1^-$ and $A_k^+/A_1^+$ (panels {\it b} and {\it c} of Fig.~\ref{fig.ampFUN}). For $A_k^-/A_1^-$ we observe more `erratic' behaviour than for $A_k^+/A_1^+$. At low orders, an increase of the amplitude ratio (above 1) is possible, just as it is observed in RR~Lyr.  $A_k^+/A_1^+$ decreases with increasing order. A plateau is visible for harmonic orders between 3 and 5.

Panel {\it d} of Fig.~\ref{fig.ampFUN} shows the variation of the $Q_k$ parameter defined as \citep{al03},
\begin{equation}
Q_k=\frac{A_k^+-A_k^-}{A_k^++A_k^-}\,,
\label{kuka}\end{equation}
 with $k$.  The line at $Q=0.3$ is marked for reference \citep[the distribution of $Q$ from the MACHO database peaks at this value,][]{al03,kol11}. It is clear that in all our models $Q_k$ is positive at low harmonic orders and hence, the amplitudes of the higher frequency triplet components are higher. In the intermediate range of harmonic orders the situation is reverted and at largest $k$ the higher frequency components are higher again.

\begin{figure}
\resizebox{\hsize}{!}{\includegraphics{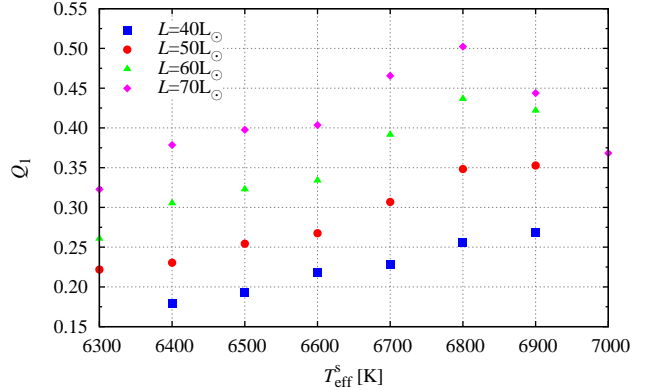}}
\caption{$Q_1$ plotted vs. the \teffs{} for models with different luminosities of the initial non-modulated models (sets M7L4, M7, M7L6, M7L7).}
\label{fig.QL}
\end{figure}

We focus our attention on the triplet components around the main frequency $f_0$. The amplitudes of these triplet components are the most robust outcome of our model and should be compared with observation first. In about 75 per cent per cent of Blazhko RR~Lyrae stars, the higher frequency side peak has a larger amplitude than the lower frequency side peak and thus $Q_1$ is positive \citep{al03}. This is in agreement with our models. However, in 25 per cent of the observed Blazhko variables $Q_1$ is negative, as the lower frequency side peak has a larger amplitude. Unfortunately, in none of our models this is the case. In Fig.~\ref{fig.QL} we plot the values of $Q_1$ vs. \teffs{} for the models of sets M7L4, M7, M7L6, M7L7, with different luminosities of the initial non-modulated models. The trend of the decreasing $Q_1$ values with decreasing luminosity as well as with decreasing temperature is clearly visible. However, in no case $Q_1$ is negative. This is true for all model sequences we have investigated, regardless of the properties of turbulent convection modulation (period, amplitude and shape). As in a quarter of known Blazhko RR~Lyrae stars $Q_1$ is negative, we regard the failure to reproduce negative $Q_1$ values in our models as another significant challenge for the Stothers mechanism.

\subsection{Period doubling phenomenon}\label{sec.PD}

The period doubling phenomenon, which manifests in alternating shapes of the light curves of the consecutive pulsation cycles, was discovered very recently in {\it Kepler} RR~Lyr data \citep{kol10a}. The effect is also clearly visible in two other {\it Kepler} targets, V808~Cyg (KIC 4484128) and V355~Lyr (KIC 7505345), both being Blazhko variables \citep{szabo10}. In some other Blazhko stars, the confirmation of period doubling requires longer observation. The strength of period doubling depends on the phase in the Blazhko cycle. Period doubling was not found in any non-modulated RR~Lyrae variable so far \citep{szabo10}.

The phenomenon of period doubling is not new. Period doubling is
clearly present in RV~Tauri stars which show alternating deep and
shallow minima in their light and radial velocity curves. It was
also found in radiative hydrodynamic models of W~Vir stars
\citep{bk87,kb88}, Cepheids and BL~Her variables
\citep{mb90,mb91,bm92}. \cite{mb90} traced the origin of the period
doubling in hydrodynamic models to the destabilising role of the
half integer resonance $(2n+1)\omega_0=2\omega_k$ between the
fundamental mode and a higher order overtone ($n$ equal to 1 or 2).
They showed that the period doubling can occur close to the
resonance centre when the resonant overtone mode is either excited
or only weakly damped. \cite{mb90} also showed that, depending on
the amount of dissipation in the model, the half-integer resonance leads
either to single period-doubling or to the period-doubling cascade
(Feigenbaum cascade) and chaos.

The period doubling phenomenon in Blazhko RR~Lyrae stars observed by
{\it Kepler} was analysed in detail by \cite{szabo10}, who also
proposed the underlying mechanism. Their convective hydrodynamic
models (with fixed convective parameters) display period doubling,
the origin of which was traced to the 9:2 resonance between the
fundamental mode, and the ninth order overtone,
$9\omega_0=2\omega_9$. In these models, the ninth overtone is a
trapped envelope mode \citep[see e.g.,][]{bk01} with a much higher
growth rate than that computed for the neighbouring overtone modes.
As noted above, the weak damping of the ninth overtone favours the
occurrence of period doubling. In their detailed analysis
\cite{kms11} used the relaxation technique \citep{stel74} to
determine the stability of the fundamental mode pulsation through
the Floquet stability coefficients. They showed that indeed the
fundamental mode is destabilized through the $9\omega_0=2\omega_9$
resonance and excluded all other possible half-integer resonances.
In addition, they showed that the $9\omega_0=2\omega_9$ resonance
can lead not only to a single period doubling, but also to the
period-four and period-eight limit cycles (the Feigenbaum cascade).

Our models also display period doubling, clearly visible in the light curves (Fig.~\ref{fig.6800time}, Figs.~\ref{fig.papa}--\ref{fig.paps}) and in the frequency spectra (Figs.~\ref{fig.6800spec} and \ref{fig.6800ampy}). It occurs only in models of higher temperatures, with \teffs{} in a range 6700\Ke--6900\Ke{} (depending on the model sequence) and always at phases around maximum pulsation amplitude and minimum values of the mixing length (see right columns of Figs.~\ref{fig.papa}--\ref{fig.paps}). Our linear analysis confirms the findings of \cite{szabo10} and \cite{kms11}. During the phases of minimum mixing length, in which period doubling occurs, the equilibrium models computed assuming corresponding, fixed values of the mixing length are very close to the $9\omega_0=2\omega_9$ resonance centre. We note that the mixing length has to be sufficiently small. Period doubling does not occur in models with small amplitude of the mixing-length modulation (Fig.~\ref{fig.papa}). Also, the ninth overtone in our models is trapped in the outer envelope and is close to being unstable.

For the period doubling to occur, the model has to be close to the
resonance condition. As discussed by \cite{szabo10}, if the Stothers
mechanism is indeed operational in Blazhko variables and the
convective structure of the star varies during the Blazhko cycle (as
it does in our models), it is natural that period doubling occurs
only at certain Blazhko phases, at which the physical conditions are
favourable. Qualitatively, this is what we see in our models.
Favourable conditions (proximity to the 9:2 resonance) occur during the
phases of low mixing length.  These are the phases at which period
doubling occurs. We note that in our models period doubling is
strictly repetitive and always appears at the same phase of the
Blazhko cycle. In fact, at other phases the period doubling does not
vanish entirely. Its remnants serve as a seed for the fast growth of
alternations during the next Blazhko cycle.
We would like to stress that turbulent convection itself does not
cause period doubling, which is a resonant phenomenon and may occur
in purely radiative models as well. The modulation of turbulent
convection in our models just changes the structure of the outer
stellar layers, bringing the instantaneous model periods close to
the resonance condition.

As discussed above, the transient occurrence of the period doubling during the Blazhko cycle can be naturally interpreted within the Stothers model. Nevertheless, a more quantitative comparison with observations reveals some discrepancies. In our models period doubling is limited to the phases of maximum pulsation amplitude. For the three {\it Kepler} Blazhko RR~Lyrae stars it is most prominent in the phases preceding the maximum pulsation amplitudes, but not only, it is also visible close to the phases of minimum pulsation amplitude \citep{szabo10}. Certainly it is present in a much wider range of Blazhko phases than it is in our models.

\begin{figure}
\resizebox{\hsize}{!}{\includegraphics{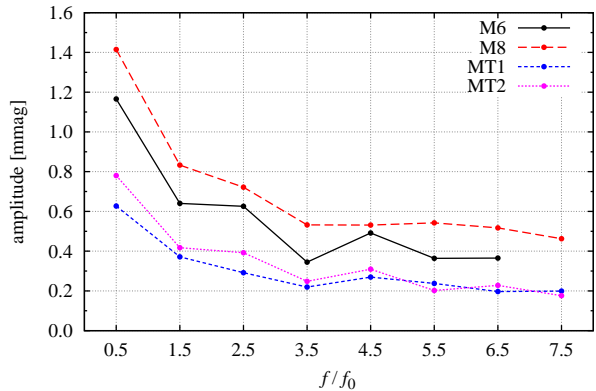}}
\caption{Amplitudes of half integer frequencies for models of sets M6, M8, MT1 and MT2 (initial non-modulated model was the same for all four models displayed, and its effective temperature was 6800\thinspace K).}
\label{fig.ampHIF}
\end{figure}

As for the frequency spectra, the amplitudes of the half integer components in all three stars observed by {\it Kepler} follow the same pattern. The highest peak is observed at $3/2\,f_0$ followed by $5/2\,f_0$ and $1/2\,f_0$. This is not what we compute in our models, as already mentioned in Section~\ref{sec.fapc}. In Fig.~\ref{fig.ampHIF} we show the amplitudes of HIF's, for models with different patterns of turbulent convection modulation (\teffs{} is the same in these models, equal to 6800\Ke). These are the same models as displayed in Fig.~\ref{fig.ampFUN}. The highest peak is always located at $1/2\,f_0$ followed by $3/2\,f_0$ and $5/2\,f_0$. For the latter two peaks the amplitudes are comparable. A local maximum is visible at $9/2\,f_0$ for our models of sets M6, MT1 and MT2. Such a maximum is expected if the 9:2 resonance is indeed responsible for the period doubling. The fact that it is very weak and not present in all our models is a puzzle.

\section{Additional models}\label{sec.discussion}

\begin{figure*}
\centering
\resizebox{1.\hsize}{!}{\includegraphics{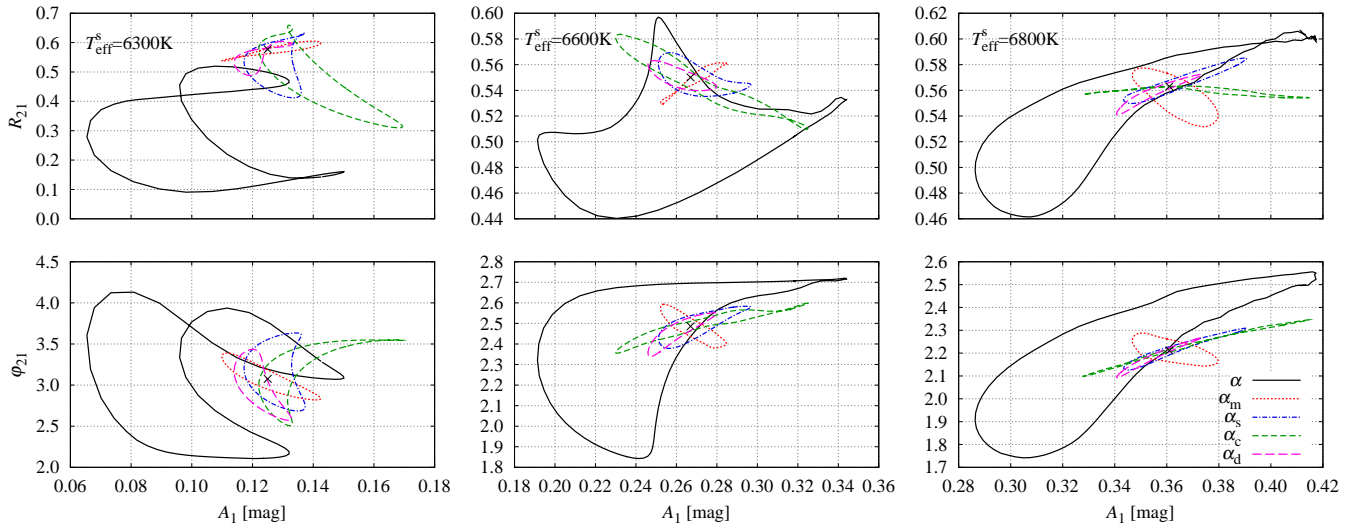}}
\caption{Light curve variation through the Blazhko cycle. $R_{21}$ vs. $A_1$ and $\varphi_{21}$ vs $A_1$ for models in which different $\alpha$-parameters of the convective model are modulated.}
\label{fig.alphas}
\end{figure*}

The results presented so far reveal some disagreements between the computed models and the observations. The frequency analysis shows that in all our models $Q_1$ is positive, so the higher frequency side peak around the fundamental mode frequency has a higher amplitude than the lower frequency side peak. In about 25 per cent of Blazhko variables $Q_1$ is negative. Also, we cannot reproduce the details of the light curve variation. In particular the anticorrelation between amplitude ratio, $R_{21}$, and amplitude, $A_1$, observed in RR~Lyr through the majority of the Blazhko cycle cannot be reproduced. Probably, the most important challenge for the Stothers mechanism is a huge range of variation of the mixing length. This large variation is required to reproduce the ranges of light curve variation in strongly modulated stars like RR~Lyr, as well as the pulsation period changes we observe in Blazhko variables.

To get an idea whether and how these difficulties may be overcome, we have computed additional sequences of models in which we vary other $\alpha$-parameters of the turbulent convection model than the mixing length (see Section~\ref{sec.code}). So far, we modulated the mixing length only, which is the natural choice, as the mixing-length parameter controls the overall strength of convection. Now, as an exercise, we vary the other parameters entering the turbulent convection model. We modulate the strength of eddy-viscous dissipation ($\alpha_{\rm m}$), the strength of the source function ($\alpha_{\rm s}$), the strength of the convective heat flux ($\alpha_{\rm c}$) and the strength of the turbulent dissipation ($\alpha_{\rm d}$, turbulent-cascade term). During the mixing-length modulation, all the above terms were modulated simultaneously, as these $\alpha$-parameters enter the model in pair with the mixing-length parameter \citep[$\alpha\alpha_{\rm m}$, $\alpha\alpha_{\rm s}$, $\alpha\alpha_{\rm c}$ and $\alpha_{\rm d}/\alpha$; see e.g.,][]{sm08a}. We note that there is no physical justification behind the modulation of any particular term listed above. Only a detailed 3D magnetohydrodynamical computation could clarify how a variable magnetic field could affect particular phenomena connected with the turbulent convection. Unfortunately, such computations and even appropriate models do not exist currently.

In all computed model sequences, the shape of modulation is sinusoidal, its period is 40\thinspace days and the amplitude of its modulation is 50 per cent. Particular $\alpha$-parameters vary around the values defined in Table~\ref{tab.convpar.nonmodulated}.

In Fig.~\ref{fig.alphas} we plot the $R_{21}$ vs. $A_1$ and $\varphi_{21}$ vs. $A_1$ relations for models with three different initial temperatures (\teffs=6300\Ke, 6600\Ke{} and 6800\Ke), just like in Figs.~\ref{fig.papa}--\ref{fig.paps}. The cross in each panel corresponds to the initial non-modulated model. Thick solid trajectories correspond to the model of set M7 in which the mixing length is modulated (default in this paper). As expected, the ranges of variation of the Fourier parameters are smaller if we vary other $\alpha$-parameters of the model than the mixing length, as in each case only one term of the convective model is modulated. We focus on $R_{21}$ vs. $A_1$ relation (upper panels of Fig.~\ref{fig.alphas}). Observed trends are not always simple (particularly for models with \teffs=6300\Ke) and they depend on the temperature of the model. The increase of $R_{21}$ with decreasing $A_1$, as we observe in RR~Lyr, can be reproduced most easily when only the strength of the convective heat flux is modulated. Then, however, the range of variation of the Fourier parameters is small, despite a rather huge modulation of the convective heat flux. Therefore, it is difficult to propose a modulation to get a better model for RR~Lyr. We have to vary more than one convective parameter to get a large range of variation of the Fourier parameters. For the model with \teffs=6600\Ke{} one could modulate only $\alpha_{\rm s}$, $\alpha_{\rm c}$ and $\alpha_{\rm d}$ and keep the eddy-viscous dissipation fixed. On the other hand, Fig.~\ref{fig.alphas} suggests that the same modulation adopted in the hotter model (\teffs=6800\Ke) would lead to increasing $R_{21}$ with increasing $A_1$.

\begin{figure}
\resizebox{\hsize}{!}{\includegraphics{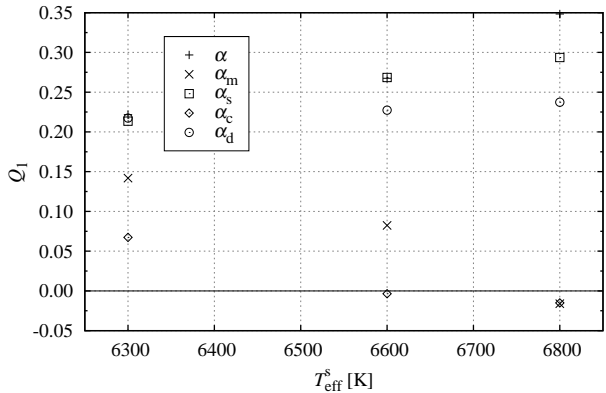}}
\caption{$Q_1$ plotted vs. the \teffs{} for models, in which particular terms of the convective model were modulated.}
\label{fig.Qa}
\end{figure}

We have also investigated whether the negative $Q_1$ values can be obtained through modulating particular terms in the convective model. In Fig.~\ref{fig.Qa} we plot the $Q_1$ vs. \teffs{} for the discussed models. Negative, albeit very close to zero, values of $Q_1$ are obtained only for the hottest models for which either only convective heat flux is modulated (\teffs=6600\Ke{} and \teffs=6800\Ke) or only eddy-viscous dissipation is modulated (\teffs=6800\Ke). Modulation of other components of the convective model always leads to positive $Q_1$.

In all our models, the convective parameters vary around the values defined in Table~\ref{tab.convpar.nonmodulated}. As noted in Section~\ref{sec.cms}, with such values we can reproduce the observational properties of the non-Blazhko RR~Lyrae variables and also classical Cepheids (with smaller eddy-viscous dissipation). With these parameters we neglect the effects of turbulent pressure and overshooting from the convective zone. The computation of models including these effects is very time-consuming. In order to check whether the inclusion of turbulent pressure and overshooting can change our results we have computed one additional model sequence in which we set $\alpha_{\rm p}=1.0$ and $\alpha_{\rm t}=0.01$. We adopted the same modulation parameters as for set M6 (Table~\ref{tab.modul}) i.e., sinusoidal modulation with a period equal to 60\thinspace d and an amplitude of $A=50$ per cent. The results are qualitatively the same as described in the previous sections. All models are characterised by positive values of $Q_1$. For the lower temperatures, the light curve variation is qualitatively the same as presented in, e.g., Fig.~\ref{fig.papa} (set M6). At higher temperatures, the range of variation of the Fourier parameters becomes smaller for models including turbulent pressure. Consequently, for such models, an even larger amplitude of turbulent convection modulation would be necessary to reproduce the strongly modulated Blazhko variables.

\section{Summary and Conclusions}\label{sec.conclusions}

In this paper we have investigated whether the mechanism proposed by \cite{st06} is capable of reproducing the light curve variation and the properties of the frequency spectra we observe in RR~Lyrae Blazhko variables. The mechanism proposed by Stothers is extremely complicated, as it assumes the coupling between a variable magnetic field and turbulent convection in high amplitude, strongly non-linear pulsators. To get a variable magnetic field a rotational/turbulent dynamo is postulated. There are no comprehensive models dealing with all these processes. Even the current models to deal with turbulent convection in non-linear pulsation are rather simple, 1D, one-equation formulas.

Recent observational progress poses serious problems for the two most popular models to explain the Blazhko effect, the Magnetic Oblique Rotator/Pulsator model and Non-radial Resonant Rotator/Pulsator model (see Introduction). The Stothers mechanism remains as a scenario that has not been confronted with concrete challenges from observations. The variable, tangled magnetic fields postulated by \cite{st06} cannot be easily detected, making the idea hard to verify on purely observational grounds \citep{kb09}. Its stochastic nature makes it attractive in the light of irregularities commonly detected in the Blazhko cycles of many variables. However, this is only a general idea, not supported by any detailed calculations or modelling. To advance with the theory we proposed a simple model to check whether the variation of convective structure of the star can lead to the modulation we observe in Blazhko stars. The modulation of turbulent convection is introduced into our models in an {\it ad hoc} way, neglecting the dynamical coupling with the postulated magnetic field.

Comparison of our models with overall properties of Blazhko
variables, as well as a detailed comparison with the strongly
modulated prototype RR~Lyr, observed by {\it Kepler}, reveals
several discrepancies with the observations and challenges for the
Stothers mechanism. In our opinion, the most important objection is
the required strength of the turbulent convection modulation. In
order to reproduce the ranges of the light curve variation observed
in strongly modulated stars like RR~Lyr, we have to modulate the
mixing-length by up to $\pm$50 per cent on a time-scale of several
tens of days. The same estimate results from the analysis of
pulsation period changes \citep[Section~\ref{sec.ppc}, see
also][]{mok10}. The physical reality of such strong modulation is,
in our opinion, questionable, although definite claims require
detailed magnetohydrodynamic modelling of the problem, which is
beyond the scope of this paper. We note here that it is not possible
to infer the mixing-length variation in the subatmospheric layers
directly from observations. In a recent paper, \citet{pre11}
analysed the variation of FWHMs (full width at half maximum) of
spectral lines in several RR~Lyrae-type stars. This parameter is a
measure of the atmospheric turbulence. Preston shows that maximum
value of FWHM varies with the Blazhko cycle. However, it is not
clear if changing intensity of the atmospheric turbulence relates in
any way to changes of the mixing-length in the subatmospheric
layers. A strong variation of the FWHM occurs also in non-modulated
RR~Lyrae stars (see Preston's fig.~3), and, we may add, also in
non-modulated hydrodynamical models with constant mixing-length
\citep[see, e.g.,][]{bs85}. In our opinion, the variation of the
maximum FWHM reflects the behaviour of the turbulence generated by
the velocity gradients in the atmosphere, rather than the possible
variation of the mixing length in the deeper layers. \citet{pre11}
shows that the maximum value of the FWHM is strongly correlated with
the pulsation amplitude, which varies during the Blazhko cycle (his
fig.~6). The larger the pulsation amplitude, the larger the maximum
value of the FWHM. On the other hand, we note that the maximum of
the FWHM occurs at pulsation phases close to the minimum radius.
These are the phases at which the velocity gradients in the
atmosphere are the strongest. Strong velocity gradients generate
strong turbulence. Thus, the larger the pulsation amplitude, the
stronger the atmospheric velocity gradients and, consequently, the
stronger the turbulence and the higher the FWHM
maximum.\footnote{The question whether variations of the FWHM in
the Blazhko stars are predominantly caused by changing velocity
gradients in the atmosphere (as we expect, and which is the case for
non-modulated stars) or whether possible modulation of the
sub-atmospheric convection can also play a role, can be examined in
more detail by numerical computations. This requires repeating the
analysis of \cite{bs85} for both the modulated models and for
non-modulated models of different pulsation amplitudes. Such work
however, is beyond the scope of this paper.}

The light variation we compute resembles that observed in Blazhko variables, however, the details cannot be reproduced, at least for RR~Lyr. We note that our results can be used in the future for comparison with other Blazhko variables for which satellite data will be soon available. This would clarify how severe the discrepancies between the model and the observed light variation are. As for the frequency spectra, we cannot reproduce the asymmetry of the modulation side peaks around the main pulsation frequency. In our models, the higher frequency side peak always has a higher amplitude, which is not the case in about a quarter of the Blazhko variables \citep{al03}.

The critical analysis of the Stothers idea by \cite{GezaSF} is also worth mentioning. He points out that the predictions by \cite{st06,st10} of expected period changes through the instability strip, that agree with values observed in some Blazhko stars, are not correct, as they result from a misinterpretation of the linear and non-linear model periods.

In the light of our results, the idea proposed by \cite{st06} faces difficulties, and should be treated with caution. It needs refinement and further detailed studies in order to put it on solid physical grounds. Certainly, it is worth further investigation, but other possibilities to explain the Blazhko phenomenon should be explored as well.

\section*{Acknowledgments}
Funding for this Discovery mission is provided by NASA's Science Mission Directorate. The authors gratefully acknowledge the entire {\it Kepler} team, whose outstanding efforts have made these results possible.

Model computations presented in this paper were conducted on the psk computer cluster in the Copernicus Centre, Warsaw, Poland. We are grateful to James Nemec for comments on this manuscript. RS and KK are supported by the Austrian Science Fund (FWF projects AP 21205-N16 and T359/P19962, respectively).


\label{lastpage}

\end{document}